\documentclass[12pt,preprint]{aastex}

\shorttitle{{\it AKARI} Near-infrared Spectroscopy of NGC~1097}
\shortauthors{Kondo et al.}

\begin{document}

\title{The Central Region of the Barred Spiral Galaxy NGC~1097 Probed by {\it AKARI} Near-Infrared Spectroscopy}

\author{Toru Kondo\altaffilmark{1}, Hidehiro Kaneda\altaffilmark{1}, Shinki Oyabu\altaffilmark{1}, Daisuke Ishihara\altaffilmark{1}, Tatsuya Mori\altaffilmark{1}, and Mitsuyoshi Yamagishi\altaffilmark{1}}
\affil{Graduate School of Science, Nagoya University}
\email{kondo@u.phys.nagoya-u.ac.jp}

\author{Takashi Onaka\altaffilmark{2} and Itsuki Sakon\altaffilmark{2}}
\affil{Graduate School of Science, The University of Tokyo}

\and

\author{Toyoaki Suzuki\altaffilmark{3}}
\affil{Institute of Space and Astronautical Science, Japan Aerospace Exploration Agency}

\altaffiltext{1}{Graduate School of Science, Nagoya University, Chikusa-ku, Nagoya 464-8602, Japan}
\altaffiltext{2}{Graduate School of Science, The University of Tokyo, Bunkyo-ku, Tokyo 113-0033, Japan}
\altaffiltext{3}{Institute of Space and Astronautical Science, Japan Aerospace Exploration Agency, Sagamihara, Kanagawa 252-5210, Japan}

\begin{abstract}
With the Infrared Camera on board {\it AKARI}, we carried out near-infrared (2.5--5.0 $\mu$m) spectroscopy of the central kiloparsec region of the barred spiral galaxy, NGC~1097, categorized as Seyfert 1 with a circumnuclear starburst ring. Our observations mapped the area of $\sim50\arcsec\times10\arcsec$ with the resolution of $\sim 5\arcsec$, covering about a half of the ring and the galactic center. As a result, we spatially resolve the starburst ring in the polycyclic aromatic hydrocarbon 3.3 $\mu$m, the aliphatic hydrocarbon 3.4--3.6 $\mu$m features, and the hydrogen Br$\alpha$ 4.05 $\mu$m emission. They exhibit spatial distributions significantly different from each other, indicating that the environments vary considerably around the ring. In particular, the aliphatic features are enhanced near the bar connecting the ring with the nucleus, where the structure of hydrocarbon grains seems to be relatively disordered. Near the center, the continuum emission and the CO/SiO absorption features are strong, which indicates that the environments inside the ring are dominated by old stellar populations. The near-infrared spectra do not show any evidence for the presence of nuclear activity. 

\end{abstract}


\section{Introduction}

The 2.5--5.0 $\mu$m near-infrared (NIR) spectroscopy of nearby galaxies is very important to investigate their interstellar environments, because the NIR region contains many pieces of information on the interstellar medium (ISM). They include hydrocarbon emissions such as the polycyclic aromatic hydrocarbon (PAH) feature at 3.3 $\mu$m and aliphatic sub-features at 3.4--3.6 $\mu$m, hydrogen recombination lines such as Br$\alpha$ at 4.05 $\mu$m, and absorption features such as those due to $\mathrm{H_2O}$ and CO around 3 $\mu$m and 4.7 $\mu$m, respectively. NIR continua are often attributed to photospheric emission of old cool stars and/or hot dust emission in active regions. With their spatial information, we can probe local conditions of the ISM within a galaxy.

NGC~1097 is a nearby barred spiral galaxy, which is categorized as Seyfert 1 with the starburst ring of $\sim$2 kpc in diameter and the primary and second bars of $\sim$15 kpc and $\sim$1 kpc in length, respectively \citep{ger88,sha93,sto93,koh03,hsi08}. Hence NGC~1097 is an ideal laboratory to study the ISM in galaxies showing both active galactic nucleus and circumnuclear starburst activities. The optical and NIR images reveal the dust lanes running along the primary bar outside the ring and the second bar inside the ring \citep{qui95,pri05}. Both bars appear to have contact points at the starburst ring. The gas flowing into the ring along the primary bar may have triggered the circumnuclear starburst, while the second bar may have driven gas into the nucleus, possibly fueling the central supermassive blackhole \citep{pri05,fat06,dav09}. \citet{hsi11} showed that the CO (J=2-1) emission is bright in the ring, the dust lanes, and at the contact points of the ring with the primary bar.

In this Letter, we present the 2.5--5.0 $\mu$m NIR spectra of NGC~1097 obtained with the Infrared Camera (IRC; \citealt{onaka07}) on board the {\it AKARI} satellite \citep{mura07}. Our observations mapped the area of $\sim 50\arcsec \times 10\arcsec$ with the resolution of $\sim 5\arcsec$, covering about a half of the ring and the galactic center. Over a large central area, we detected the PAH 3.3 $\mu$m feature, 3.4--3.6 $\mu$m sub-features, Br$\alpha$ emission, and absorption features due to $\mathrm{H_2O}$ and CO/SiO. On the basis of the spectra, we discuss the interstellar environment of the central region of NGC~1097. We below adopt the distance of 19.1 Mpc to NGC~1097 \citep{will97}, for which a projected scale of $1\arcsec$ corresponds to 92 pc.

\section{Observations and Data Reduction}

The NIR spectroscopic observations of the central region of NGC~1097 were performed within the framework of the {\it AKARI} mission program ``ISM in our Galaxy and Nearby galaxies" (ISMGN; \citealt{kane09}) in the {\it AKARI} post-helium phase (phase 3). The two pointed observations (ObsIDs 1420462.1 and 1420462.2) were carried out on July 18, 2008. To obtain the 2.5--5.0 $\mu$m spectra, we used a grism spectroscopic mode (R $\sim$ 120) with the slit of $5\arcsec\times48\arcsec$ for its width and length, respectively \citep{oh07}. Figure 1 shows the slit positions of the two observations and the slit sub-aperture regions, from which we created the spectra in Fig.~2. The second slit position (slit B) is shifted by $\sim4\arcsec$ from the first slit (slit A), as shown in Fig.~1.

The basic spectral analyses were performed by using the standard IDL pipeline prepared for reducing phase 3 data with the newly calibrated spectral response curve.\footnote{\url{http://www.ir.isas.jaxa.jp/ASTRO-F/Observation/}} We adopted a slit sub-aperture with the length of $4\farcs5$ (3 pixels), and created a local spectrum for every slit sub-aperture. We determined the length of the slit sub-aperture so that it could give as high spatial resolution as possible with sufficiently high S/N ratios. Considering that the FWHM of the point spread function of the IRC is $\sim$3 pixels \citep{onaka08}, we adopted the 3 pixel length for the slit sub-aperture. We applied smoothing to each spectrum with a boxcar kernel of 3 pixels ($\sim$0.03 $\mu$m).

\section{Result}

Most of the spectra in Fig.~2 show strong PAH emission at 3.3 $\mu$m with sub-features at 3.4--3.6 $\mu$m and Br$\alpha$ emission at 4.05 $\mu$m. Some also exhibit broad absorption features around 3 $\mu$m and/or 4.5 $\mu$m. Here we consider that the absorption at $\sim 3\,\mu\rm{m}$ is attributed to interstellar $\mathrm{H_2O}$ ice (3.05 $\mu$m) by judging from its spatial distribution, as described below. The absorptions at $\sim 4.5\,\mu\rm{m}$ are probably due to the blending of SiO (4.30 $\mu$m) and CO (4.66 $\mu$m) gas in the photospheres of old O-rich stars, such as K and M stars \citep{coh92, her02}. The continuum brightness is highest at A6, which is closest to the galactic center, monotonically decreasing with distance from the center. All the spectra have negative slopes against the wavelength, suggesting that cool stars of old populations make a significant contribution to the continuum emission.

In order to quantify the overall continuum spectra, we fit the spectral regions at 2.55--2.70 $\mu$m and 3.60--4.00 $\mu$m by a power-law function. After subtraction of the best-fit power-law continuum, we fit the PAH feature and the 3.41 $\mu$m sub-feature by Lorentzian profiles and the 3.46, 3.51, and 3.56 $\mu$m sub-features and the Br$\alpha$ emission by Gaussian profiles \citep{mori12}, as well as the absorption features due to $\mathrm{H_2O}$, SiO, and CO by Lorentzian profiles. We choose Lorentzian and Gaussian profiles for resolved broad features and unresolved emission lines, respectively. The central wavelength of each profile is adjusted and fixed by considering the redshift of NGC~1097. The widths of the PAH feature, the 3.41 $\mu$m sub-feature, and the $\mathrm{H_2O}$ absorption feature are fixed at the best-fit values obtained from the spectrum that has the highest S/N ratio for each component. The widths of the other sub-features and Br$\alpha$ are fixed at 0.034 $\mu$m, the spectral resolution of the IRC, while those of the SiO and CO absorption features are set to be free. We consider the errors given by the pipeline plus the systematic errors originating in the continuum; the latter errors are evaluated by changing the aforementioned continuum fitting regions by $\pm$0.05 $\mu$m. The red solid curves in Fig.~2 indicate the best-fit model thus obtained for the continuum plus the lines and features of each spectrum.

Figure 3 shows the spatial variations of the fluxes of the spectral components along the slits, which are calculated by integrating the flux densities over the corresponding wavelength ranges not only for the spectra in Fig.~2 but also for those from other slit sub-apertures of the same size at intermediate positions. We shift the slit sub-aperture position by 1 pixel one after the other to derive local spectra consecutively along the slit in Fig.~3. The equivalent widths of the absorption features due to $\mathrm{H_2O}$, SiO, and CO are derived by dividing their absorption fluxes by the best-fit power-law continuum flux densities at the central wavelengths. The continuum brightness at 3.3 $\mu$m and the index derived by the power-law fitting to the continuum spectrum are also shown in Fig.~3. As can be seen in the figure, the spatial profiles are categorized into three different types; the PAH, sub-features, Br$\alpha$ emissions, and the power-law index exhibit double-peaked structures, which appear to be associated with the starburst ring. The CO plus SiO absorption feature and the 3.3 $\mu$m continuum features
  are broadly distributed with a peak toward the galactic center, while the $\mathrm{H_2O}$ absorption feature is not apparently associated with either ring or center.

In order to better visualize the spatial distribution of each spectral component, we show the two-dimensional spectral maps in Fig.~4, overlaid on the contour map of the {\it Spitzer}/IRAC 3.6 $\mu$m band. The spectral maps were created by regridding the discrete data points in Fig.~3 to the spatial bin of $1\farcs7\times1\farcs7$ on the equatorial coordinate plane. The figure clearly shows that the PAH, sub-features, and Br$\alpha$ emissions are associated with the starburst ring, while the others are not.

\section{Discussion}

We spatially resolve the starburst ring in the PAH emission feature by spectroscopy for the first time. We also resolve the ring in the Br$\alpha$ emission and the 3.4--3.6 $\mu$m sub-features. They exhibit spatial distributions significantly different from each other; the PAH 3.3 $\mu$m emission shows a relatively uniform distribution as compared with the other two components, which is similar to the far-infrared (FIR) distribution of large grains obtained by the {\it Herschel}/PACS \citep{sand10}. The Br$\alpha$ emission shows a local peak in the eastern part of the ring; since Br$\alpha$ is an indicator of star-formation activity \citep{ima10}, the starburst activity is likely to be enhanced locally in this region as also suggested by FIR spectroscopy with the {\it Herschel}/PACS \citep{bei10}. The uniform distribution of the PAHs then indicates that they are not dominantly heated by young stars in such active local regions, but rather by relatively old stars, which are distributed widely in the central region as shown below. The result agrees with the idea suggested by \citet{sand10} for stars later than B type as dominant heating sources of large grains. The 3.4--3.6 $\mu$m sub-features are locally enhanced toward this active region, and more interestingly, also the inside of the ring spatially corresponding to part of the second bar that connects the ring with the center (Fig.~1). The latter is more evident in the ratio map of the sub-features to the PAH 3.3 $\mu$m feature intensity in Fig.~5. The 3.4--3.6 $\mu$m sub-features are probably attributed to small carbonaceous grains with aliphatic structures \citep{dul81,kwo11}. Thus the local enhancements of the aliphatic features suggest that small hydrocarbon grains, not much aromatized, may be newly formed through shattering of carbonaceous grains \citep{jon96,kane12,yama12}, which provide observational evidence that the gas and dust in the second bar is really in a turbulent motion, streaming into the nucleus from the ring.

The $\mathrm{H_2O}$ ice absorption feature has a spatial distribution entirely different from the CO and SiO absorption features, neither of which are likely to be associated with the starburst ring. The former feature shows local maxima at positions similar to those in the CO (J=2--1) emission \citep{hsi11}, which is reasonable because the $\mathrm{H_2O}$ ice is likely to be formed in such dense gas regions. In particular, the enhancement is located at the CO clump near the contact point of the primary bar with the ring, where the CO gas has relatively low excitation temperatures but high velocity dispersion \citep{hsi11}. The result suggests that an ice formation mechanism may work efficiently in such a condition. On the other hand, the SiO and CO absorption features are similar in spatial distribution to the 3.3 $\mu$m continuum emission, which is reasonable because both probably originate from photospheres of old stars. Hence old stellar populations are dominant over the central region. There is no contribution of hot dust at the center, because the CO/SiO equivalent width does not decrease toward the center, and also because the power-law index of the continuum does not change at the center, either (Fig.~3). There seems to be no starburst activity at the center, since Br$\alpha$ is not significantly detected. Hence we cannot find any evidence for the presence of nuclear activity from our observations. In order to reconcile this result with the previous detection of a central compact source at 11.7 $\mu$m and 18.3 $\mu$m \citep{mas07}, we suggest that the dust in the nucleus is rather cool so that it cannot contribute to the continuum emission at wavelengths at shorter than 5.0 $\mu$m. We estimate the dust temperature to be lower than 250 K, referring to 48 mJy at 11.7 $\mu$m for the nuclear emission \citep{mas07} with the condition of $<3\,{\rm MJy\,sr^{-1}}$ at 5.0 $\mu$m from our result ($<10\,\%$ of the observed flux density). In contrast, the CO/SiO equivalent width decreases and the power-law index increases in the circumnuclear ring, both of which imply the presence of hot dust and thus intense starburst activity in the ring.

\section{Summary}

We presented the result of the near-infrared (2.5--5.0 $\mu$m) spectroscopy of central $\sim$1 kpc area of NGC~1097 with the {\it AKARI}/IRC. From all the regions of the starburst ring, we detected the PAH 3.3 $\mu$m feature and the aliphatic 3.4--3.6 $\mu$m sub-features as well as the Br$\alpha$ 4.05 $\mu$m emission. They exhibit spatial distributions significantly different from each other, indicating that local conditions of the ISM vary much from region to region around the ring. The PAH 3.3 $\mu$m emission shows a relatively uniform distribution as compared with the other two components, which indicates that they are not dominantly heated by young stars in active regions of the ring, but rather by old stars distributed widely over the central region. In particular, the aliphatic features are enhanced near the bar connecting the ring with the nucleus, where the structure of hydrocarbon grains seems to be relatively disordered. This suggests that the gas may be flowing into
  the nucleus along the bar in a turbulent motion, which causes the shattering of carbonaceous grains in shocks. Near the galactic center, however, these emissions are considerably suppressed, while the 3.3 $\mu$m continuum and the CO/SiO absorption features are strong. Because the latter are thought to originate from cool old stars, we conclude that the environments inside the ring are dominated by old stellar populations. We find no evidence for the presence of an active nucleus from our observations.

\acknowledgments

We thank all the members of the {\it AKARI} project. {\it AKARI} is JAXA project with the participation of ESA. This research is supported by a Grant-in-Aid for Scientific Research No. 22340043 from the Japan Society for the Promotion of Science, and the Nagoya University Global COE Program, ``Quest for Fundamental Principles in the Universe: from Particles to the Solar System and the Cosmos'', from the Ministry of Education, Culture, Sports, Science and Technology of Japan.

\clearpage

\begin{figure}
\epsscale{}
\begin{center}
\includegraphics[width=0.7\linewidth]{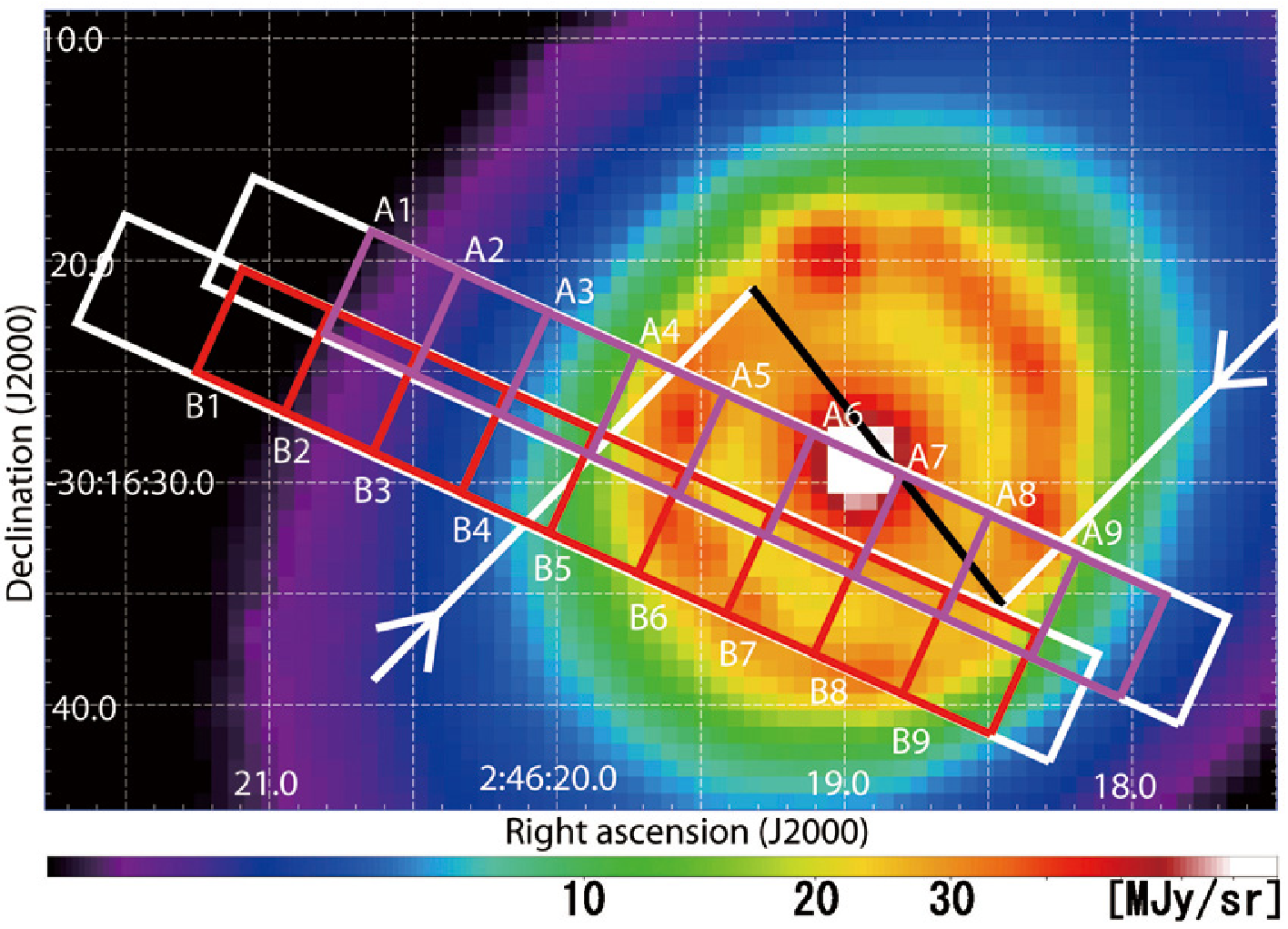}
\end{center}
\caption{Positions of the two slit apertures (white boxes) and their sub-apertures (magenta and red boxes) used to create the spectra in Fig.~2, overlaid on the {\it Spitzer}/IRAC 3.6 $\mu$m image of the central region of NGC~1097. Each slit sub-aperture (A1--A9, B1--B9) has a size of $5\arcsec\times4\farcs5$. The approximate locations of the primary and the second bar are indicated by the white and the black line, respectively \citep{qui95,pri05}. The arrows indicate the expected motion of the gas in the primary bar.}
\end{figure}

\begin{figure}
\epsscale{}
\plotone{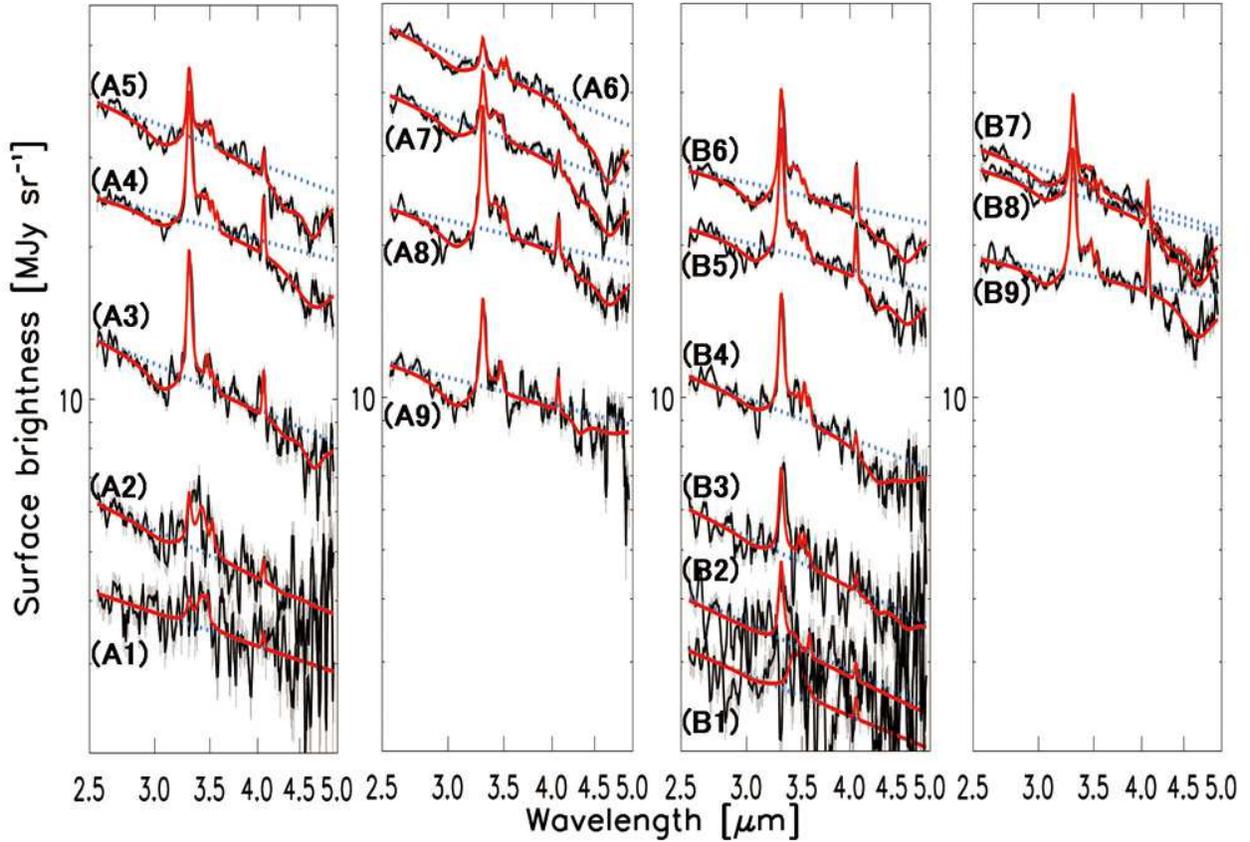}
\caption{{\it AKARI}/IRC 2.5--5.0 $\mu$m spectra together with the names of the slit sub-apertures indicated in Fig.~1. The left two and the right two panels show the spectra observed in slits A and B, respectively. For each spectrum, the red solid curve indicates the best-fit model, where the continuum is fitted by a power-law, while the lines and features are by Lorentzian or Gaussian profiles (see text for details). The blue dotted curve indicates the best-fit continuum.}
\end{figure}

\begin{figure}
\plotone{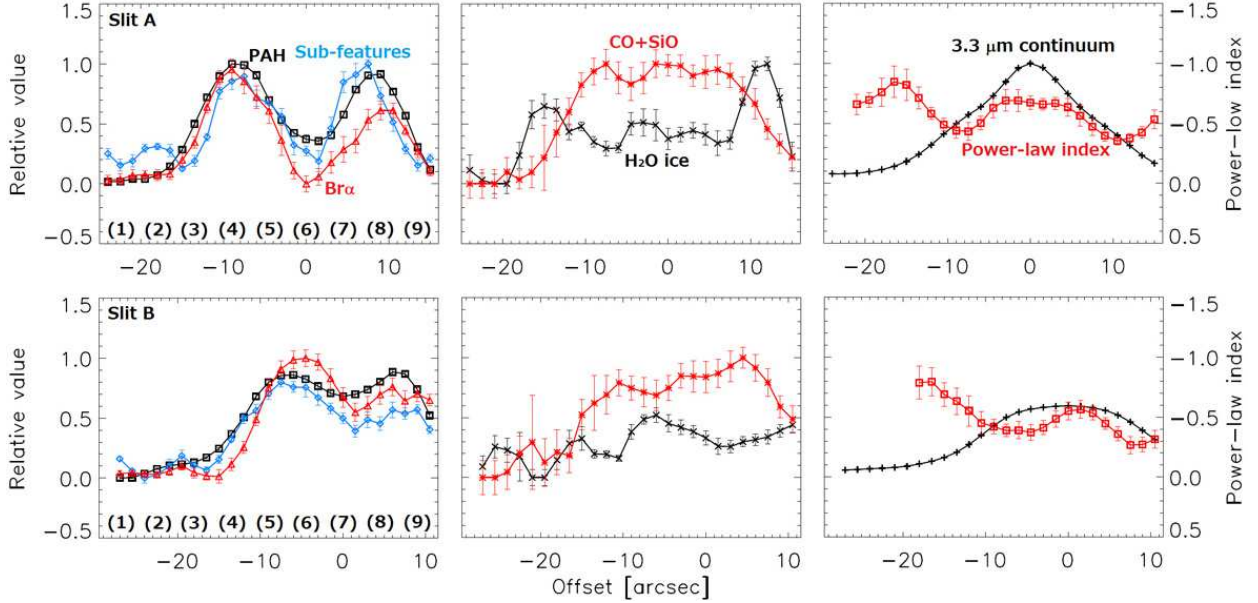}
\caption{Changes of the spectral components with the offsets from A6 to A1$\sim$A9 (top) and from B7 to B1$\sim$B9 (bottom). The squares, diamonds, and triangles in the left panels correspond to the intensities of the PAH 3.3 $\mu$m feature (peak: $3.7\times10^{-4}\,{\rm erg\,s^{-1}\,cm^{-2}\,sr^{-1}}$), 3.4--3.6 $\mu$m sub-features ($1.7\times10^{-4}\,{\rm erg\,s^{-1}\,cm^{-2}\,sr^{-1}}$), and Br$\alpha$ ($3.9\times10^{-5}\,{\rm erg\,s^{-1}\,cm^{-2}\,sr^{-1}}$), respectively. The crosses and asterisks in the middle are the equivalent widths of the $\mathrm{H_2O}$ ice (120 nm) and CO+SiO gas absorptions (207 nm), while the pluses and squares in the right are the continuum brightness at 3.3 $\mu$m ($45\,{\rm MJy\,sr^{-1}}$) and the index derived by the power-law fitting to the continuum. The power-law indices are shown only for the regions where the continuum is twenty times higher than the background level at 5 $\mu$m. Each profile is normalized to unity at a peak except for 
 that of the power-law index.}
\end{figure}

\begin{figure}
\begin{center}
\begin{minipage}{.32\linewidth}
\includegraphics[width=1.0\linewidth]{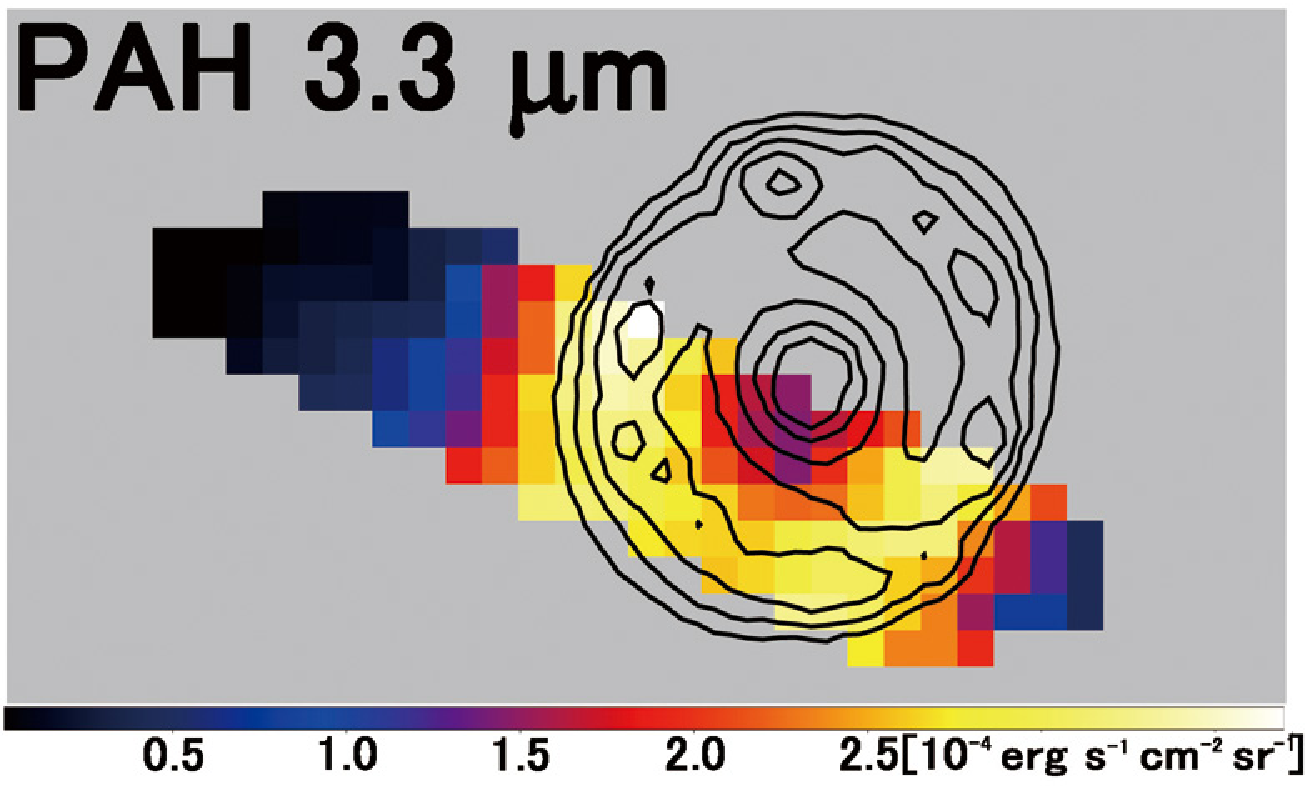}
\end{minipage}
\begin{minipage}{.32\linewidth}
\includegraphics[width=1.0\linewidth]{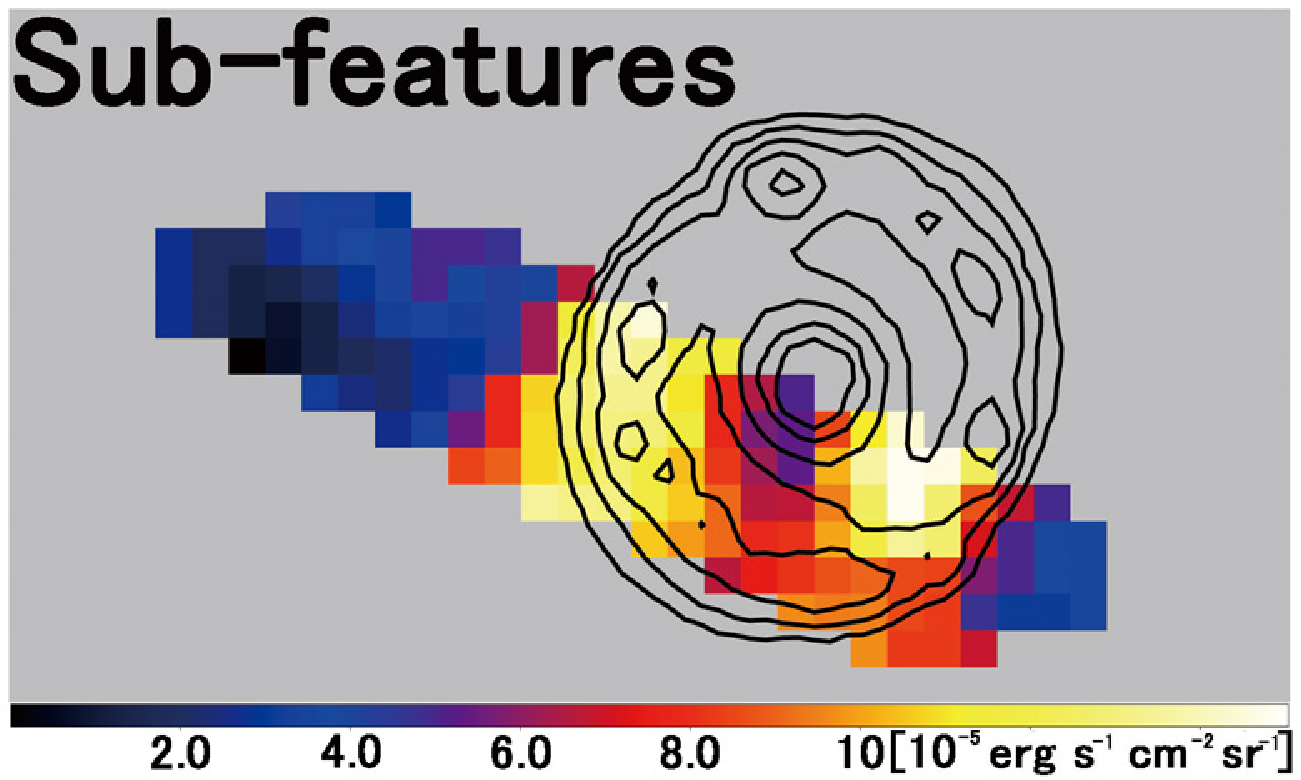}
\end{minipage}
\begin{minipage}{.32\linewidth}
\includegraphics[width=1.0\linewidth]{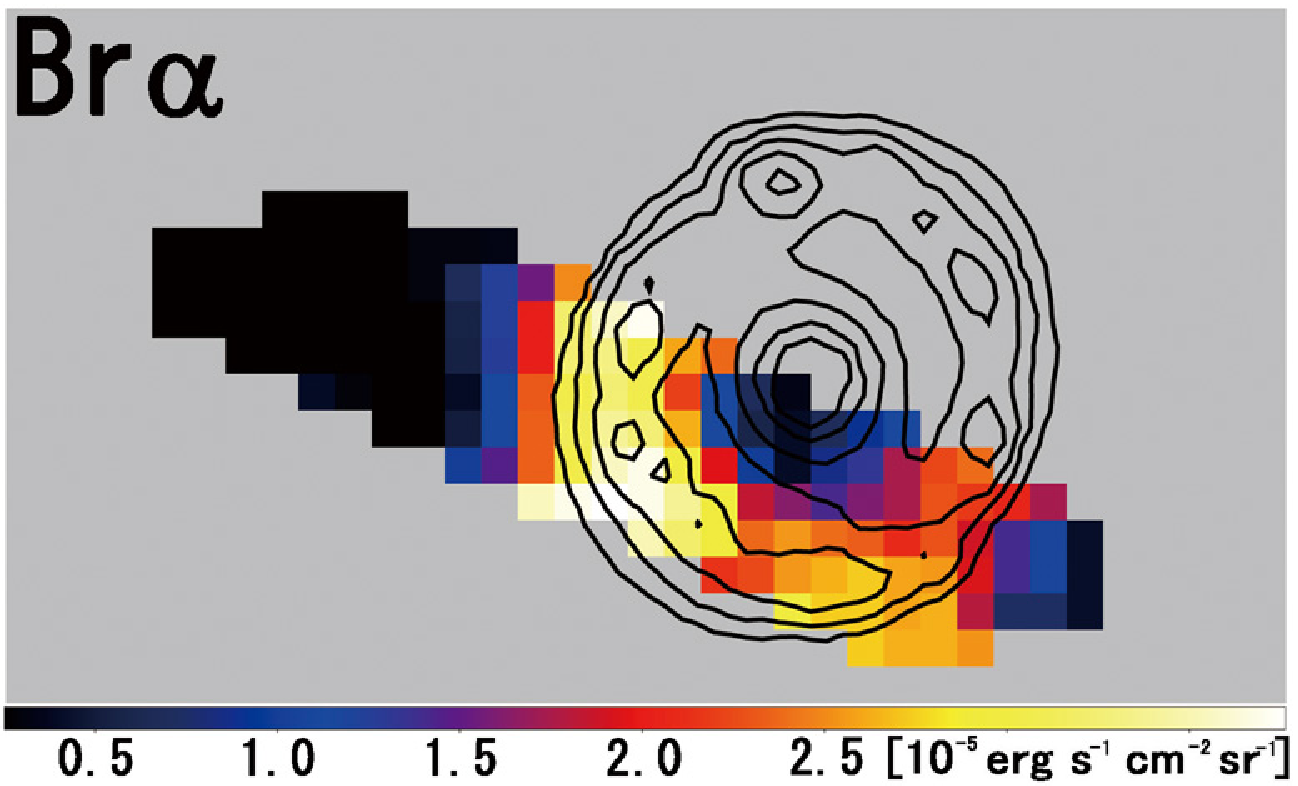}
\end{minipage}

\begin{minipage}{.32\linewidth}
\includegraphics[width=1.0\linewidth]{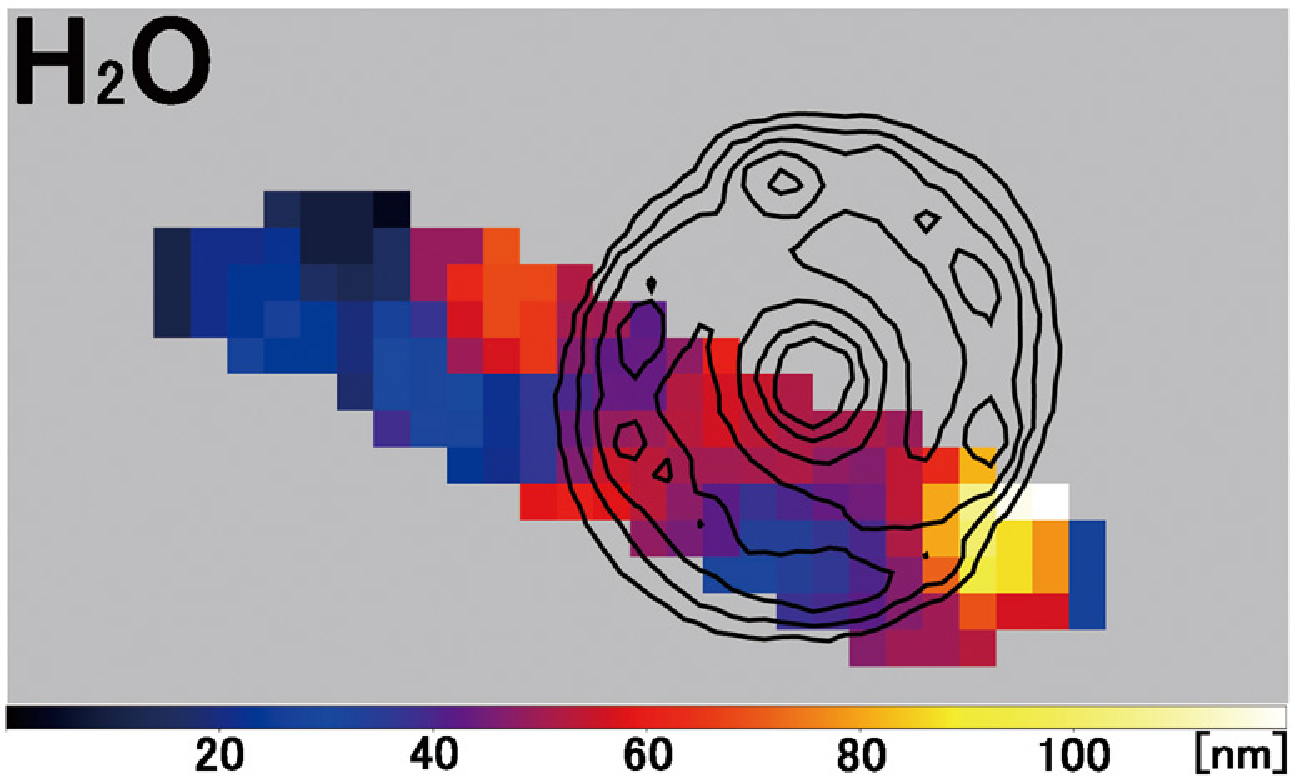}
\end{minipage}
\begin{minipage}{.32\linewidth}
\includegraphics[width=1.0\linewidth]{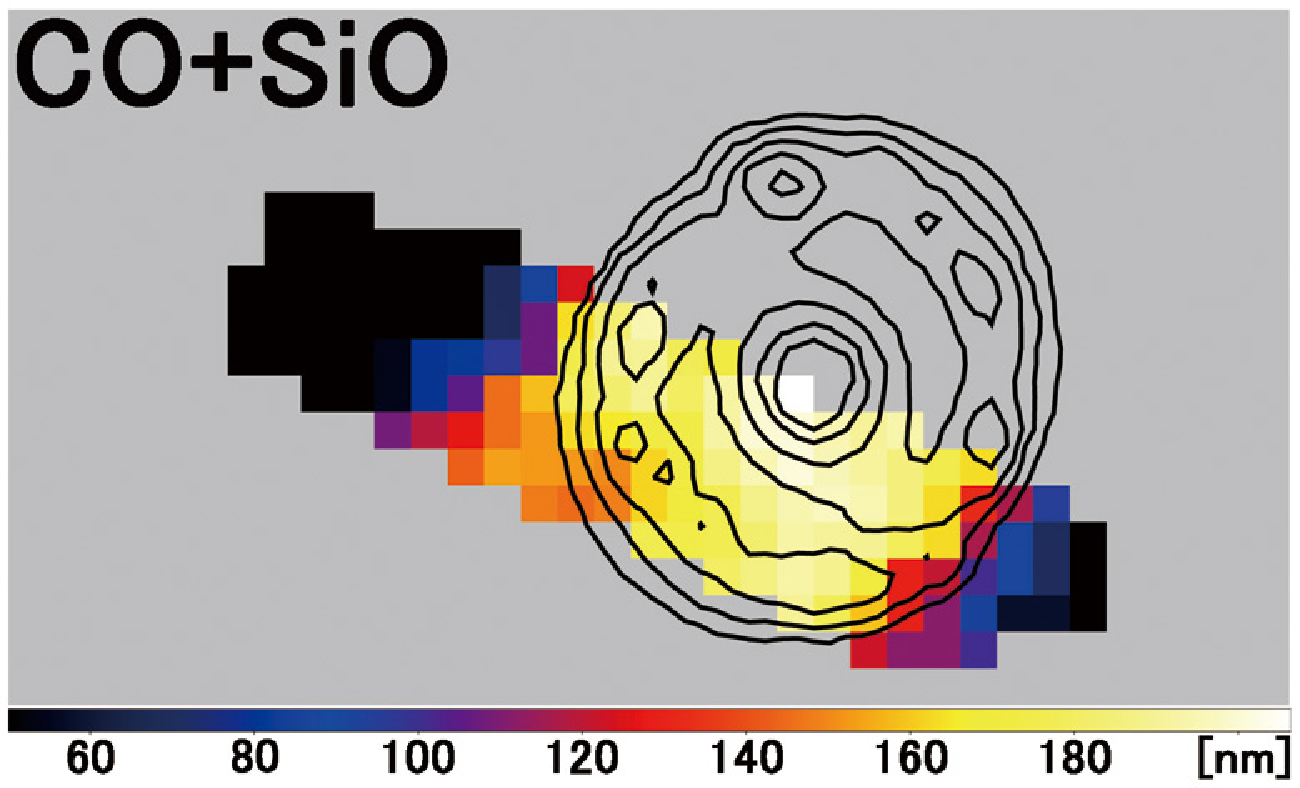}
\end{minipage}
\begin{minipage}{.32\linewidth}
\includegraphics[width=1.0\linewidth]{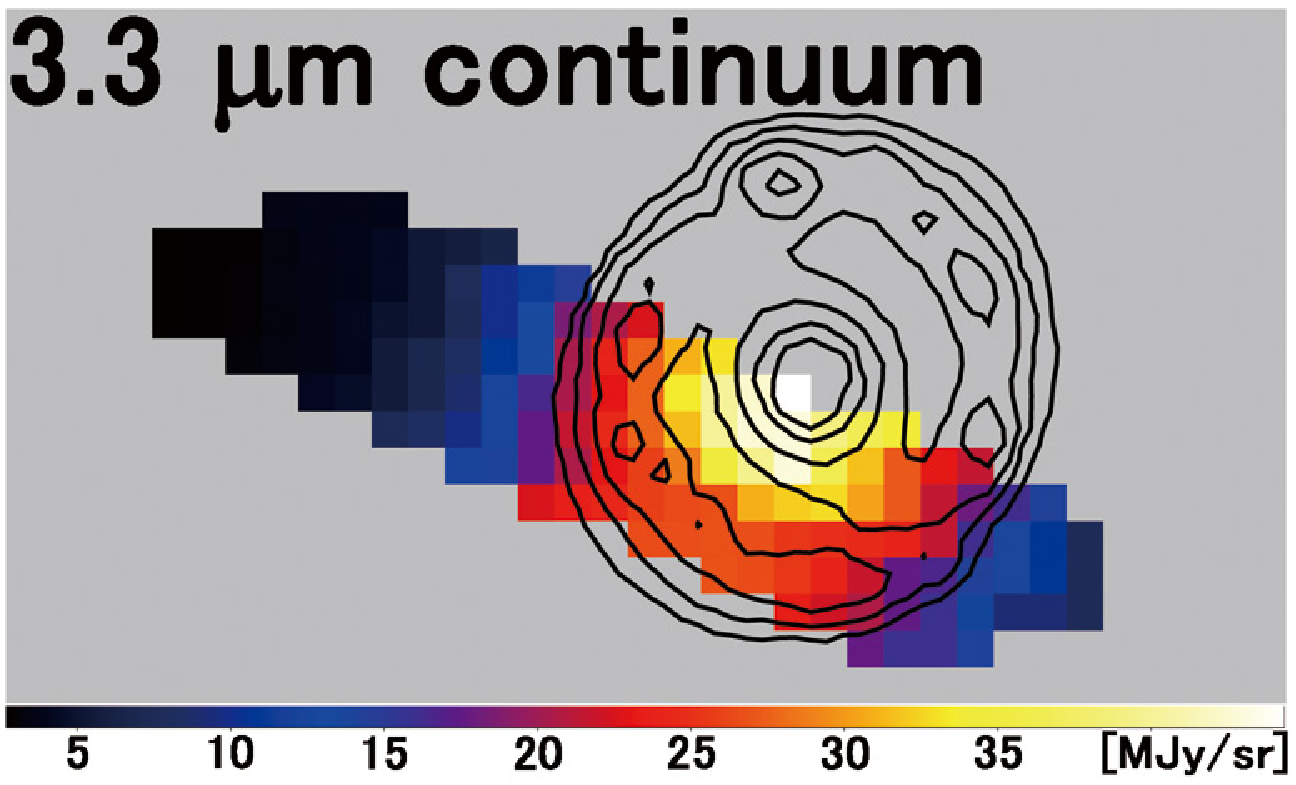}
\end{minipage}
\end{center}
\caption{Spectral maps of the PAH 3.3 $\mu$m feature, 3.4--3.6 $\mu$m sub-features, Br$\alpha$, $\mathrm{H_2O}$ ice absorption, CO+SiO gas absorption, and the continuum brightness at 3.3 $\mu$m, overlaid on the contour map of the {\it Spitzer}/IRAC 3.6 $\mu$m band (Fig.~1). The contours are drawn at logarithmically-spaced six levels from 50\% to 15\% of the peak. The absorption features are measured in the equivalent widths. We adopt a linear color scale from the minimum to the maximum for each map.}
\end{figure}

\begin{figure}
\begin{center}
\includegraphics[width=0.7\linewidth]{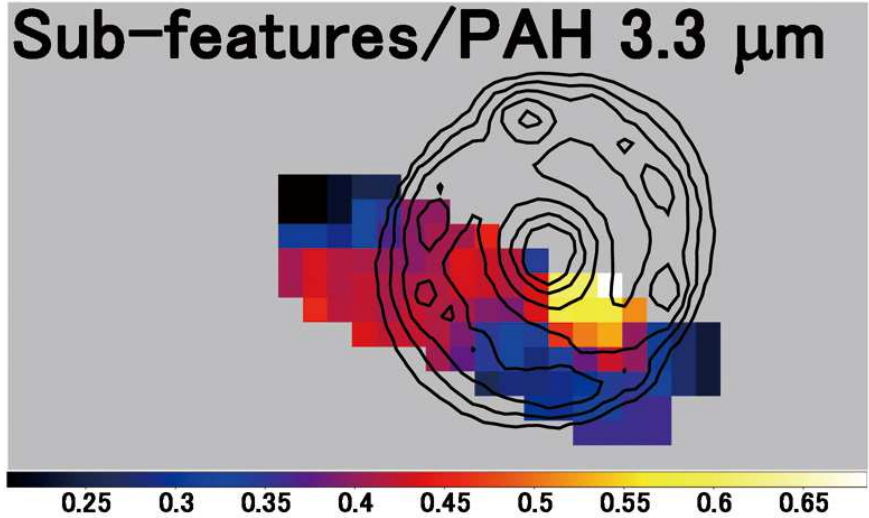}
\end{center}
\caption{Ratio map of the sub-features to the PAH 3.3 $\mu$m feature intensity, which is shown only for the region where the signal-to-noise ratios are higher than five for both PAH 3.3 $\mu$m and sub-features. The contours are the same as in Fig.~4.}
\end{figure}

\end{document}